\documentclass{raa_twocolumn} 

\usepackage{graphicx,times}         
\usepackage{natbib}
\usepackage{amssymb,amsmath}
\bibpunct{(}{)}{;}{a}{}{,}
\usepackage[pagebackref=true]{hyperref}
\usepackage{comment}

\begin{document}

\title{The \textit{SVOM} mission }
\subtitle{Its profile and its system}

\volnopage{Vol.0 (202x) No.0, 000--000} 
\setcounter{page}{1}   

\author{B. Cordier\inst{1,*}\footnotetext{$*$Corresponding Authors}
   \and J.Y. Wei\inst{2}
   \and S.N. Zhang\inst{3} 
   \and S. Basa\inst{4}
   \and J.-L.Atteia\inst{5}
   \and A. Claret\inst{1}
   \and A. Coleiro\inst{6}
   \and F. Daigne\inst{7}
   \and N. Dagoneau\inst{1}
   \and J.S. Deng\inst{2}
   \and Y.W. Dong\inst{3}
   \and O. Godet\inst{5}
   \and D. G\"otz\inst{1}
   \and X.H. Han\inst{2}
   \and C. Lachaud\inst{6}
   \and E. W. Liang \inst{8}
   \and F. Piron \inst{9}
   \and Y.L. Qiu\inst{2}
   \and S. Schanne\inst{1}
   \and D. Turpin\inst{1}
   \and S.D. Vergani\inst{10}
   \and J. Wang\inst{2}
   \and C. Wu\inst{2}
   \and L. P. Xin\inst{2}
   \and B. Zhang\inst{11}
   \and M. Bai\inst{12}
   \and S. Crepaldi\inst{13}
   \and K. Feng\inst{15}
   \and F. Gonzalez\inst{13}
   \and M. Huang\inst{2}
   \and D. Li\inst{14}
   \and Y. Liu\inst{12}
   \and H. Louvin\inst{1}
   \and K. Mercier\inst{13}
   \and J. Jaubert\inst{13}
   \and R. Su\inst{14}
   \and M.Y. Wei\inst{12}
   \and X.F. Zhang\inst{14}
   \and Y. Zhang\inst{14}
   \and the SVOM collaboration\inst{16}
}
   

\institute{
    CEA Paris-Saclay, Institut de Recherche sur les lois Fondamentales de l'Univers, 9111 Gif sur Yvette, France; {\it bertrand.cordier@cea.fr}\\
    \and National Astronomical Observatories, Chinese Academy of Sciences, Beijing 100101, P. R. China;\\
    \and Key Laboratory of Particle Astrophysics, Institute of High Energy Physics, Chinese Academy of Sciences, Beijing 100049, P. R. China;\\ 
    \and Aix Marseille Universit\'e, CNRS, CNES, LAM, Marseille, France;\\ 
    \and IRAP, Universit\'e de Toulouse, CNRS, CNES, Toulouse, France;\\ 
    \and Université Paris Cité, CNRS, CEA, Astroparticule et Cosmologie, 75013 Paris, France;\\ 
    \and Sorbonne Universit\'e, CNRS, UMR 7095, Institut d'Astrophysique de Paris, 75014 Paris, France;\\ 
    \and Guanxi Key Laboratory for Relativitic Astrophysics, Guangxi University, Nanning 530004, P. R. China;\\ 
    \and Laboratoire Univers et Particules de Montpellier, Universit\'e Montpellier, CNRS/IN2P3, 34095 Montpellier, France;\\ 
    \and LUX, Observatoire de Paris, Universit\'e PSL, Sorbonne Universit\'e, CNRS, 92190 Meudon, France;\\ 
    \and Hong Kong Institute for Astronomy and Astrophysics, University of Hong Kong, P. R. China; \\ 
    \and National Space Science Center, Chinese Academy of Sciences, Beijing 100190, P. R. China;\\ 
    \and Centre National d’Etudes Spatiales, Centre Spatial de Toulouse, Toulouse Cedex 9, France;\\ 
    \and Key Lab for Satellite Digitalization  Technology, Innovation Academy for Microsatellites, Chinese Academy of Sciences, Shanghai 201304, P. R. China; \\ 
    \and   Aerospace Information Research Institute, Chinese Academy of Sciences, P. R. China;\\ 
    \and   https://fsc.svom.org/home/collaboration/collaborators\\  
\vs\no
   {\small Received 202x month day; accepted 202x month day}
}
  
\abstract{ The \textit{SVOM} (Space-based Variable
Objects Monitor) mission, launched into low Earth orbit on 22 June 2024, is a French-Chinese multi-wavelength observatory dedicated to the study of the transient sky. Inspired by the \textit{Neil Gehrels Swift Observatory}, it consists of an autonomous rapid-slewing satellite, linked in real time to several ground-based telescopes. The space segment comprises two X-ray/gamma-ray wide-field instruments (ECLAIRs and GRM) with real-time triggering capabilities combined with two narrow-field telescopes in X-ray (MXT) and in visible (VT). In addition, the \textit{SVOM} collaboration has also developed a unique visible and NIR ground-based follow-up system to promptly respond to the gamma-ray transients detected on board. The core program of  \textit{SVOM} will provide new insights into the Gamma-Ray Burst physics by providing a homogeneous dataset covering both the prompt and afterglow emissions, as well as better studying the low luminosity and soft Gamma-Ray Burst populations. As a versatile satellite platform with fast slewing capabilities, \textit{SVOM} also comprises a Target of Opportunity program and a General Program consisting in pointed observations scheduled over the year that will both significantly contribute to the multi-messenger and time-domain astronomy.
}

\authorrunning{B. Cordier, J. Wei et al. }    
\titlerunning{The \textit{SVOM} mission}  

\maketitle


\section{Time-domain Astrophysics : the Discovery Space After Swift}           
\label{sect:intro}

Astronomy is undergoing a revolution fostered by our ability to
monitor the time-variable sky continuously, not only over large regions of the electromagnetic spectrum, but also with non-photonic messengers like neutrinos and gravitational waves. 
This context results in a renewed interest for the transient multi-messenger emission of explosive phenomena, like supernovae, gamma-ray bursts (GRB), blazars or tidal disruption events. 

Contrary to classical astronomical sources, most high-energy transients, especially GRBs, are short-lived phenomena, which can only be observed during the minutes following the explosion, up to a few days. It is therefore necessary to ensure a continuous flow of observations across all wavelengths in order to better characterize them. 
Furthermore, the study of high-energy transients requires the development of high-energy instruments deployed in space. In this respect the Neil Gehrels \textit{Swift} Observatory \citep{Gehrels+etal+2004} has proven highly successful. It has localized thousands of GRBs, which have led to the measurement of hundreds of redshifts up to z $\ge 9$ \citep{Cucchiara+etal+2011}. These observations have permitted to realize the power of GRBs as probes of the early Universe and tracers of the formation history and evolution of massive stars and the birth of stellar mass black holes. They have evidenced the existence of two broad GRB families with distinct connections to their host galaxies \citep{Zhang+etal+2009}: those due to collapsing massive stars (collapsars or type II), and those originating from mergers of compact objects (mergers or type I). The detection of hundreds of early afterglows and their transition with the prompt emission have disclosed their X-ray phenomenology and revealed key properties of GRB emitting jets. For an updated historical review of gamma-ray burst detection, see \cite{ViglianoLongo+2024}.

Despite these significant advances, several key questions remain open. These include the role of gamma-ray bursts (GRBs) as potential multi-messenger sources of gravitational waves and neutrinos, the nature of their central engines, either magnetars or black holes, and their connection to stellar explosions, as well as the study of very distant GRBs originating from the first generations of stars.

In this scientific context, the development of the \textit{SVOM} mission was driven by the ambition to ensure a continuous supply of well-localized explosive transients at a time when unprecedented observational capabilities are available for their study, and to provide the most comprehensive characterization of the detected events.
This required developing a full suite of space-borne and ground-based instruments, the tools to characterize their response, and the procedures to operate a complex science mission able to react quickly to the most unpredictable events.  

\section{\textit{SVOM} mission}
\label{sect:mission}

\begin{figure*}
\centering
\includegraphics[scale=0.15]{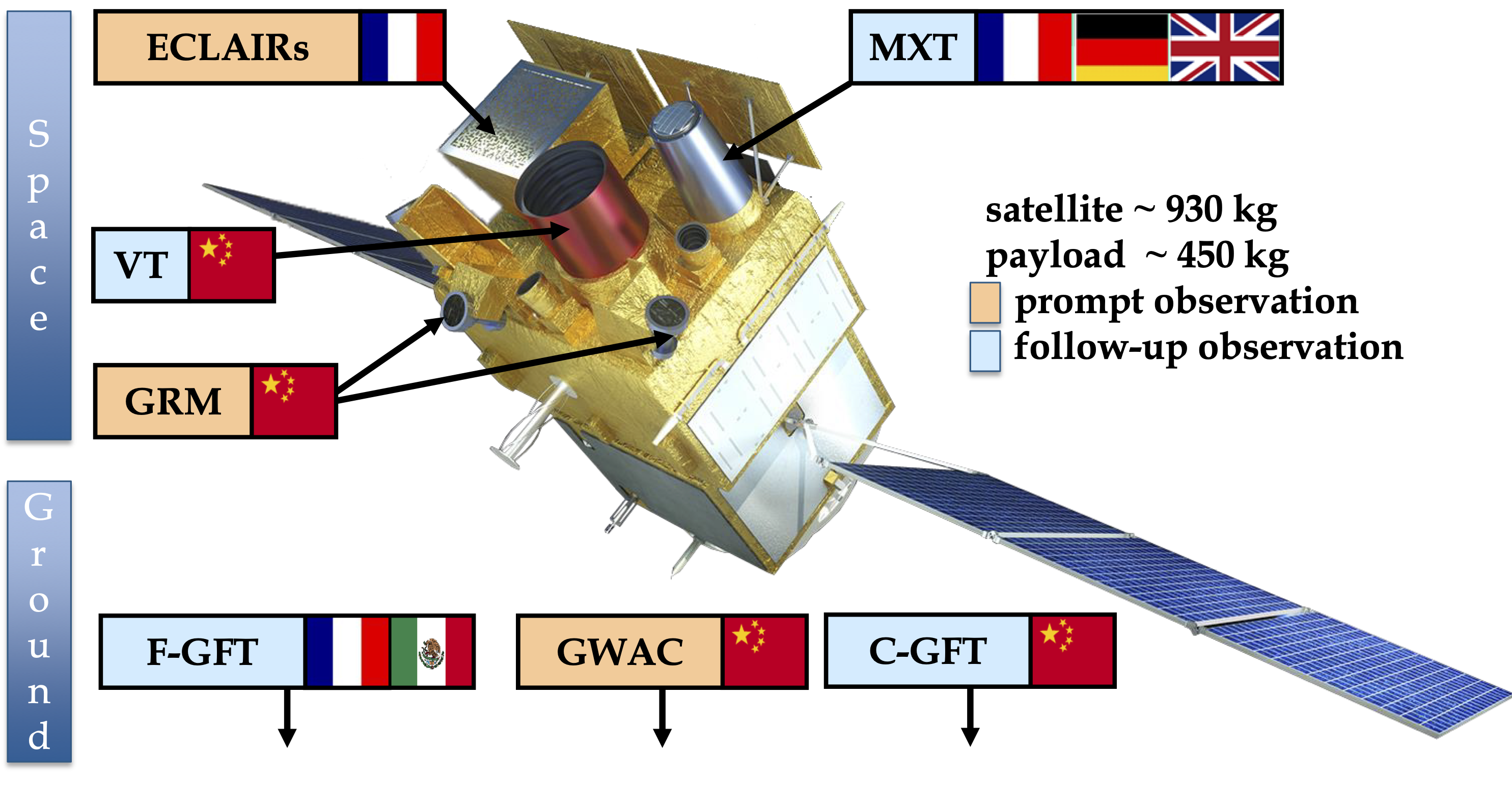}
\caption {\textit{SVOM} space-based and ground-based instruments.}
\label{fig:svom_instr}
\end{figure*}

To meet these scientific objectives, the \textit{SVOM} mission has been designed around an innovative concept: part of the payload is deployed in space, while a complementary segment is installed on the ground. The mission therefore consists of an agile satellite embarking four main instruments, which are linked via a radio channel to ground-based telescopes, see Figure~\ref{fig:svom_instr}.
The \textit{SVOM} spacecraft carries two wide-field instruments, ECLAIRs and the Gamma-Ray Monitor (GRM) for the observation of the prompt emission. ECLAIRs \citep{Godet+etal+2026a} is a coded-mask hard X-ray/gamma-ray telescope designed to detect and localize autonomously gamma-ray bursts with an precision of about ten arcminutes. Its originality lies in its energy range: it has a low-energy threshold of just 4 keV. It is complemented by the Gamma-Ray Monitor \citep{Sun+etal+2026}, which is capable of detecting gamma-ray bursts up to 5 MeV, thereby extending the spectral coverage of the prompt emission toward higher energies.
Once the burst has been localized by ECLAIRs, the satellite slews to place the source within the field of view of the narrow-field instruments: the Micro-channel X-ray Telescope (MXT, \citealt{Goetz+etal+2026}) for X-ray afterglow observations, and the Visible Telescope (VT, \citealt{Qiu+etal+2026}) for optical follow-up. In addition, MXT is designed to refine the burst localization on-board, reducing the error box to a few tens of arcseconds. The main characteristics of the \textit{SVOM} on board instruments are summarised in the Table~\ref{Tab1}.

\textit{SVOM} has also two sets of ground dedicated-instruments: a wide-field instrument GWAC (Ground-based Wide Angle Camera array) for the observation of the optical prompt emission, and two narrow-field robotic telescopes for the follow-up observations in the visible band (hereafter GFTs, Ground Follow-up Telescopes). The GFTs provide support to \textit{SVOM} by attempting to observe the early afterglow phase and to identify the optical counterpart as quickly as possible. In addition to these two dedicated telescopes, the \textit{SVOM} collaboration has access to the LCOGT network (Las Cumbres Observatory Global Telescope network), which ensures homogeneous longitudinal coverage. It is worth noting that the French-Mexican robotic telescope will be equipped with an infrared channel, expected to be operational by mid-2026. The main characteristic of the \textit{SVOM} ground-based instruments are presented in Table~\ref{Tab2} and are described in detail in \cite{Xin+etal+2026} for the GWAC and in \cite{Wu+etal+2026} and \cite{Basa+etal+2026} for the GFTs.

\begin{table*}[ht!]
\centering
\label{tab:space}
\caption{Main Characteristics of Space Instruments.}\label{Tab1}
{\begin{tabular}{l l l}
\noalign{\smallskip}\hline\noalign{\smallskip}
 & \textbf{GRM} & \textbf{ECLAIRs} \\
\noalign{\smallskip}\hline\noalign{\smallskip}
Energy range  &  15 -- 5000 keV &  4 -- 150 keV\\
Number of detectors & 3  & 6400 \\
Detection area & 3 $\times 200$ cm$^{2}$ & 950 cm$^{2}$ \\
Field of View & 2.6 sr (all 3 modules) & total: 2.0 sr ; half-coded: 1.0 sr \\
Localisation precision  & $\sim 5^\circ$ (with 3 GRDs) & 11.5'  \\
GRB rate & $\sim$120 yr$^{-1}$ 
& $\sim$60 yr$^{-1}$ \\
Additional features & Includes a particle monitor & Mask open fraction: 40\% \\
\noalign{\smallskip}\hline\noalign{\smallskip}
 &  \textbf{MXT} &  \textbf{VT}  \\
\noalign{\smallskip}\hline\noalign{\smallskip}
Energy range  &  0.2 -- 10 keV &  400 -- 650, 650 -- 1000 nm \\
Focal length &  1150 mm &  3600 mm \\
Optics dimension & 200 mm (side) &  440 mm (diameter) \\
Number of detectors & 1 (Si pnCCD) & 2 (2k $\times$ 2k CCD) \\
Effective area & 35 cm$^{2}$  @ 1.5 keV &    \\
Sensitivity (3 $\sigma$) & 10 mCrab in 10 s ; 150 $\mu$Crab in 10 ks & m$_{\mathrm V}$ = 22.5 (300 s) \\
Field of View & $58' \times 58'$ & $26' \times 26'$ \\
Localisation precision & $< 2'$ & $0.5''$ \\
Additional features & Micro-pores optics with square 40 $\mu$m pores. & Covers the full ECLAIRs \\
 & 100 ms time resolution. & error boxes. \\
\end{tabular}}
\end{table*}

\begin{table*}[ht!]
\centering
\label{tab:ground}
\caption[]{Main Characteristics of Ground-based Instruments.}\label{Tab2}
{\begin{tabular}{l l l l}
\noalign{\smallskip}\hline\noalign{\smallskip}
 &  \textbf{C-GFT} &  \textbf{C-GFT} & \textbf{FM-GFT}  \\
 &  \textit {Primary focus} & \textit{Cassegrain focus} & \textit{(aka COLIBRI)} \\
\noalign{\smallskip}\hline\noalign{\smallskip}
Location & Jilin (China) && San Pedro M\'artir (Mexico) \\
Diameter  & 1200 mm & 1200 mm & 1300 mm \\
Number of channels & 3 (g ; r ; i)& 1(g;r;i) & 3 (B/g/r/i ; z/y ; J/H) \\
Field of View  & $21' \times 21'$ & $1.28^\circ \times 1.28^\circ$ & $26' \times 26'$ (Bgrizy) ; $22' \times 22'$ (JH) \\
Limiting magnitude & m$_{AB} \approx 19$ & m$_{AB} \approx 20$ & m$_{AB} \approx 22$  \\
(r channel, 300~s, 10$\sigma$) &  &   \\
\noalign{\smallskip}\hline\noalign{\smallskip}
 &  \textbf{GWAC} & \\
\noalign{\smallskip}\hline\noalign{\smallskip}
Location & \multicolumn{2}{l}{Xinglong (China)}  \\
Number of units & \multicolumn{2}{l}{10} \\
Number of telescopes / unit & \multicolumn{2}{l}{Joint Field of View (JFoV): 4 -- Full Field of View (FFoV): 1} \\
Diameter & \multicolumn{2}{l}{JFoV: 180 mm -- FFoV: 50 mm} \\
Detector  & \multicolumn{2}{l}{JFoV: 4k x 4k CMOS -- FFoV: 6k x 4k CMOS} \\
Field of View & \multicolumn{2}{l}{JFoV: $9.5^\circ \times 9.5^\circ$ -- FFoV: $30^\circ \times 30^\circ$} \\
Limiting magnitude & \multicolumn{2}{l}{JFoV: m$_{\mathrm V} \sim 16$ -- FFoV: m$_{\mathrm V} \sim 12$ (r channel, 10~s)} \\
Additional features & \multicolumn{2}{l}{Self triggering capability ; The FFoV performs guiding.} \\
\end{tabular}}
\end{table*}

\section{The Mission Profile}
\label{sect:mission_profile}
The \textit{SVOM} satellite has been launched on 22 June 2024 by a Chinese launcher LM-2C from Xichang. Immediately after launch,  \textit{SVOM} entered the Launch Early Orbit Phase (LEOP). This phase lasted approximately 30 days and enabled us to activate and test all of the platform's systems and the on-board scientific instruments. The LEOP was followed by the in-flight commissioning phase, during which the payload was configured, and its first calibration initiated. The commissioning phase lasted 4 months. Subsequently, the validation phase was conducted to consolidate and verify the commissioning parameters, as well as to complete the calibration procedures. This validation phase extended over another period of 4 months, after which, in spring 2025, \textit{SVOM} entered its nominal scientific operational phase.

\subsection{Orbit Characteristics and Consequences }

\textit{SVOM} has been inserted into a Low Earth Orbit with an inclination of 30$^o$, an altitude of 625 km and an orbital period of $\sim$96 min.

The choice of orbit parameters  is the result of a compromise between several factors related to the overall system, where a balance must be found between technical feasibility, cost, and scientific objectives. In the case of \textit{SVOM}, the most critical factors were:
\begin{itemize}
    \item The location of the launch site in Xichang: the geographical position (latitude) of the launch pad determined the orbital inclination.
    \item The selected Long March launcher: the orbit had to remain compatible with the launcher’s capabilities.
    \item The satellite’s attitude control system, which uses the Earth’s magnetic field to desaturate the reaction wheels, is less efficient at low inclinations.
    \item The impact of charged particles on the scientific instruments, which determined the orbital altitude in order to minimize the time spent in the radiation belts.
\end{itemize}

\subsubsection{The communication channels}

\begin{figure*}
\centering
\includegraphics[width=\textwidth, angle=0]{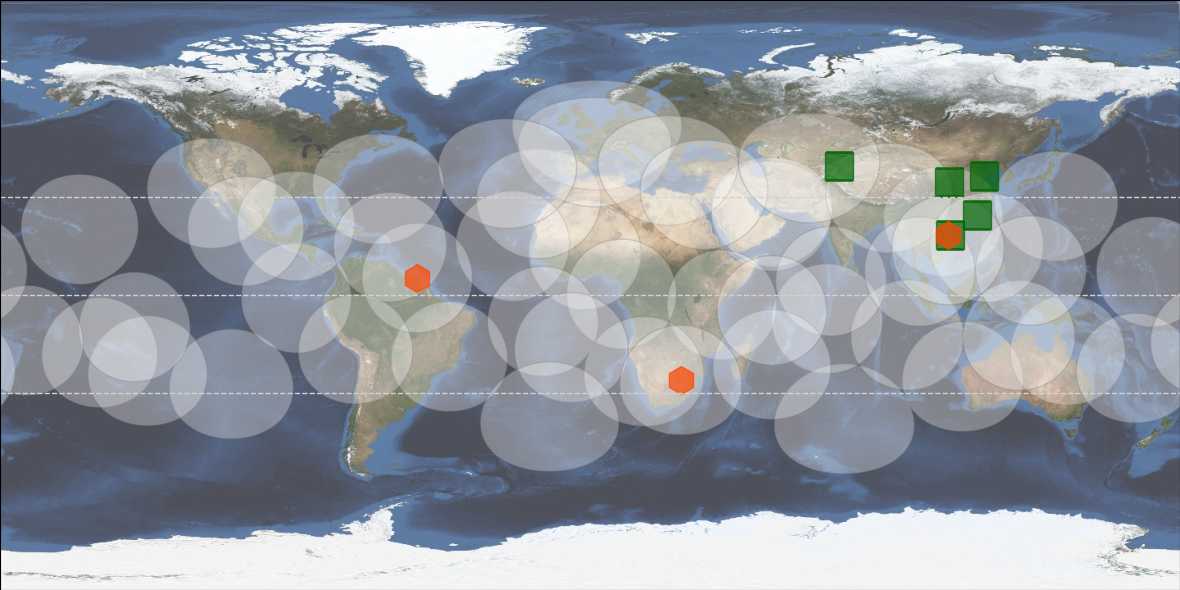}
   \caption{\textit{SVOM} communication is composed by X-band stations (orange hexagons), S-band stations (green squares) and VHF stations (white disks that are drawn from a satellite elevation of 10° and a satellite altitude of 650 km). The dotted lines represent, top to bottom, latitude +30°, the equator, and latitude -30°.}
   \label{fig:network}
   \end{figure*}

The orbital parameters of \textit{SVOM}, have a direct impact on its communication architecture. Since the mission operates in low Earth orbit, multiple ground stations distributed worldwide are required to maintain regular contact with the spacecraft, see Figure~\ref{fig:network}.

\textit{SVOM} relies on three communication channels: an X-band link dedicated to the downlink of scientific data, an S-band link for satellite command and control and a Very High Frequency (VHF) link designed to ensure the fastest possible transmission of scientific alerts.

The X-band network currently consists of 3 stations, including two French stations and one station in China. Stations included in the French network are Kourou (KUX) and Hartebeesthoek (HBX), respectively located in French Guyana and South Africa. Chinese station is located in Sanya city (Hainan, China). For operations, 8 X-band passes are planned each day: 4 passes are booked on French stations, while the other 4 are planned on Sanya station. Due to their geographical distribution, the interval between two successive X-band downlinks can reach up to 5 hours. Due to operational constraints, this time gap can be slightly longer. Efforts are ongoing to increase the number of French stations in order to shorten this latency and enable faster retrieval of the full telemetry stream.

The S-band network is composed of 5 stations, all located in China. For operations, 4 S-band passes are planned each day. In case of large amount of telecommands, extra passes can be booked. Nevertheless, the interval between two S-band passes can reach up to 12 hours and this visibility gap can be particularly critical for the rapid reprogramming of the satellite in response to exceptional targets of opportunity. To mitigate this limitation, the Beidou inter-satellite messaging system can be used, reducing the command latency to only a few minutes \citep{Bai+etal+2026}.

The VHF alert network consists of 50 stations deployed along the satellite ground track between –30° and +30° latitude. A detailed description of this system is provided in \cite{Cordier+etal+2026b}. Despite the extensive deployment, coverage gaps remain, particularly over oceanic regions, resulting in a global coverage of 93\%. The median time delay between the onboard transmission of the alert packet and the completion of data reception at the French Science Center (FSC) is approximately 9\,s (5\,s / 215\,s for the 25th and 75th percentiles), which comfortably meets the mission’s scientific requirements.

The satellite in-orbit operations and data reception are coordinated by the \textit{SVOM} Ground Support System. For a detailed description, see \cite{Liu+etal+2026}.

\subsubsection{The space environment}

Due to the tilt and offset of the Earth’s magnetic field, the inner Van Allen radiation belt approaches its minimum altitude over the South Atlantic Ocean. This region, known as the South Atlantic Anomaly (SAA), is characterized by an enhanced flux of energetic charged particles. At altitudes above 500 km, spacecrafts traversing the SAA are subject to continuous particle impacts. The approximate center of the anomaly is located near 30°S, 30°W. The extent and boundaries of the SAA are not static but vary over time as a function of solar activity.

Due  to its orbital parameters, the \textit{SVOM} satellite may intersect the SAA up to eight consecutive times per day. Within this region, the instruments are subjected to enhanced fluxes of energetic charged particles, predominantly protons and electrons. Such conditions lead to saturation of the high-energy detectors, rendering them temporarily inoperative. Consequently, the SAA imposes periods of dead time during which high-energy observations cannot be performed, thereby impacting the effective duty cycle of the mission.

\begin{figure*}
\centering
\includegraphics[width=\textwidth, angle=0]{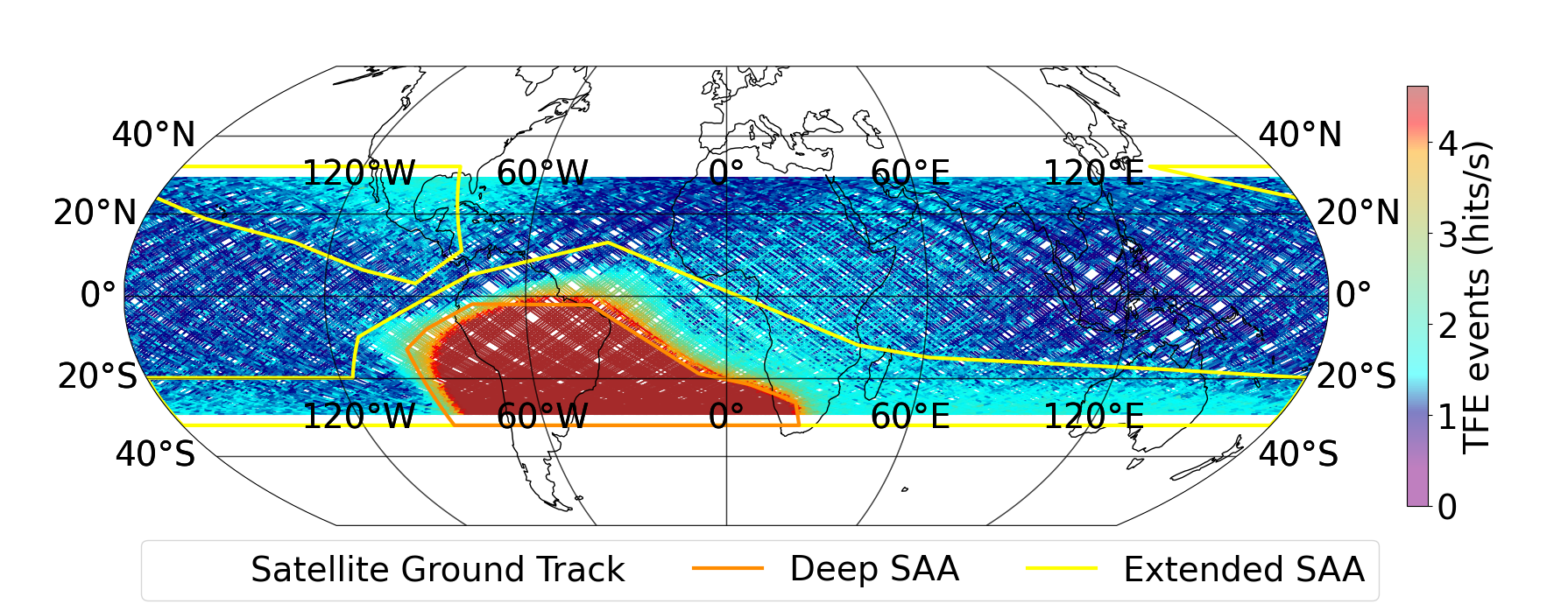}
   \caption{Rate of charge particles measured by the GRM particle counter along the orbit (arbitrary units). The yellow line delineates the extended SAA region where the satellite experiences enhanced particle fluxes, which we attribute to electrons from the outer radiation belt.}
   \label{fig:coverage}
   \end{figure*}

We expected to encounter regions of trapped particles (protons and electrons) during the passage through the SAA only. However, we also observe high count rates around latitudes of approximately –30° over the Indian Ocean and above Australia, as well as +30° over the North Pacific Ocean, see Figure~\ref{fig:coverage} which shows the map of charged particle rate measured by the GRM particle counter. These particles appear to be trapped electrons from the outer radiation belt, whose spatial extent has increased as a result of enhanced solar activity.

Indeed, although the solar maximum seems to have been reached during the second half of 2024, we remain in the maximum phase of the solar cycle, which may persist for several more months before entering the declining phase.

As a consequence of this exceptional solar activity, the availability of \textit{SVOM}’s high-energy instruments is significantly affected. For instance, in the case of the ECLAIRs instrument, the current scientific operational availability is estimated at $\sim 60\%$ \citep{Godet+etal+2026a}.

\subsection{Pointing Strategy - The Attitude Law}
In order to facilitate redshift measurements of the GRBs detected by \textit{SVOM}, the satellite attitude follows a predefined orientation, called B1 attitude law. Most of the year the optical axis of the \textit{SVOM} instruments points 45$^o$ offset from the anti-solar direction. This pointing includes avoidance periods in order to exclude the Sco X-1 source and the galactic plane from the ECLAIRs field of view. An additional constraint favors areas of the sky observable by large telescopes located in Chile, Hawa\"i and Canary Island. This strategy ensures that \textit{SVOM} GRBs are detected towards the night hemisphere, quickly observable from ground by large telescopes, and optimizes the chances to detect the GRB counterparts and host galaxies. More details on the \textit{SVOM} pointing strategy can be found in \cite{Cordier+etal+2008}.

\begin{figure}[h]
    \centering
    \includegraphics[width=\columnwidth, angle=0]{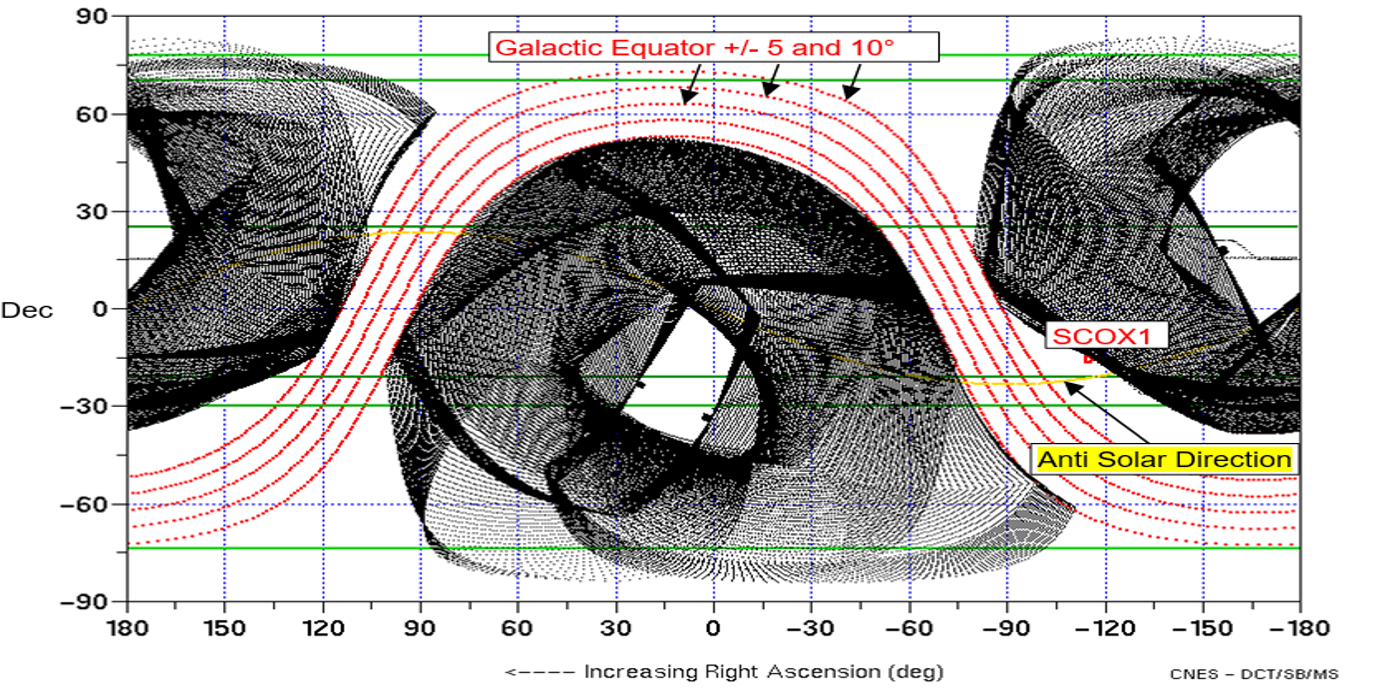}
   \caption{The ECLAIRs field of view all along the year when the satellite is pointing the Reference Attitude Law called B1. The figure is represented in equatorial coordinates. }
   \label{fig:B1_law}
\end{figure}

\begin{figure}
    \centering
    \includegraphics[width=\columnwidth, angle=0]{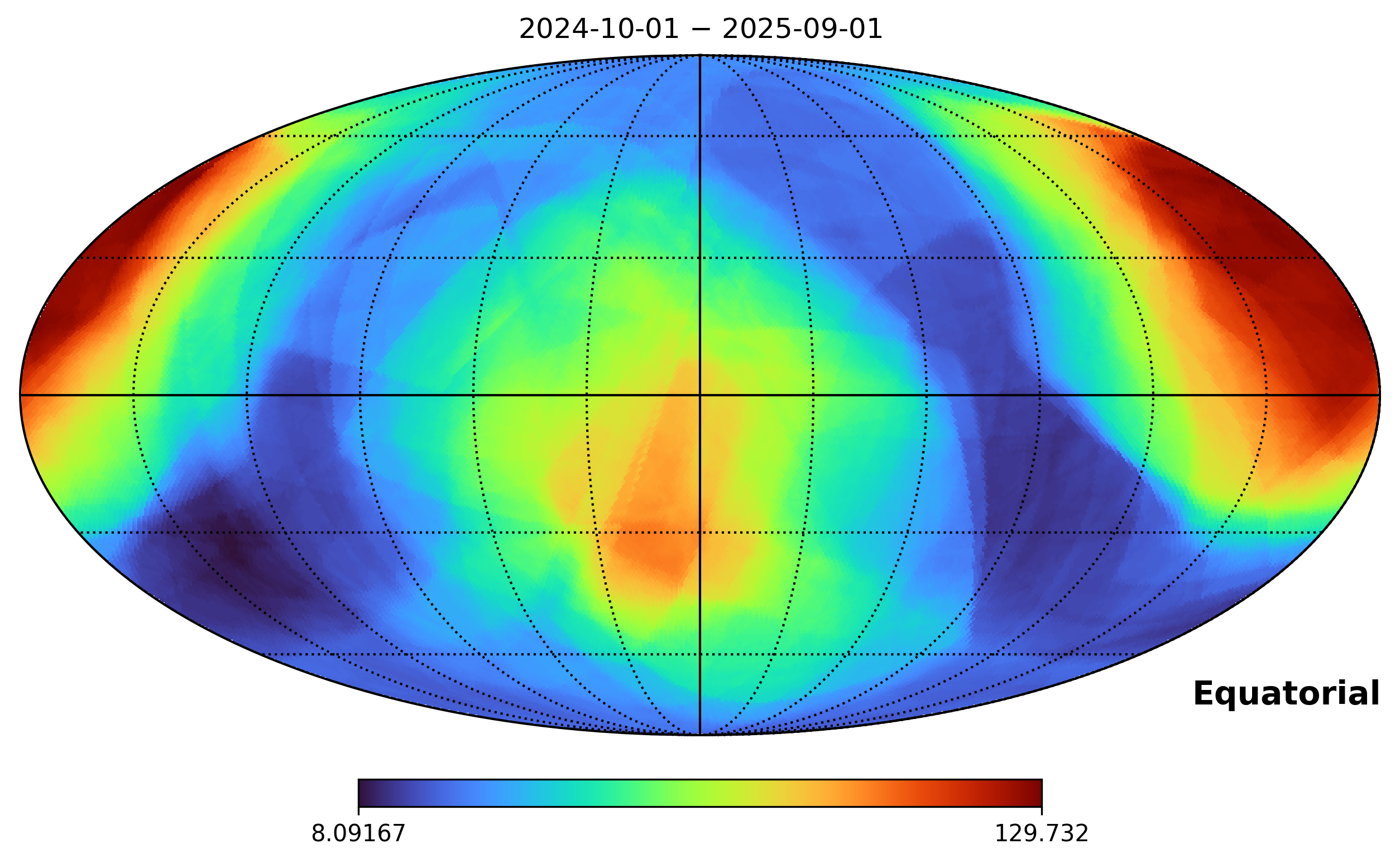}
   \caption{ECLAIRs effective exposure map of the sky in days (from 2024/10/01 to 2025/09/01, equatorial coordinates). It does not take into account the non uniform efficiency within the ECLAIRs field of view.}
   \label{fig:map_exp}
\end{figure}

As a consequence of the Low Earth Orbit combined with a roughly anti-solar attitude law, at each orbit, the Earth occults  the FoV of the \textit{SVOM} instruments. 
On average, about 65\% of the ECLAIRs field of view is accessible over one orbit (i.e., not occulted by the Earth), compared to about 50\% for the narrow-field instruments MXT and VT.

\section{The \textit{SVOM} scientific Programs}
\label{Scientific_Program}

The scientific operations of the \textit{SVOM} mission are organized into three main programs: the Core Program (CP), the General Program (GP), and the Target of Opportunity Program (ToO).

The Core Program consists of observations triggered by ECLAIRs alerts. This program focuses on gamma-ray bursts (GRBs) and all high-energy transient sources that activate the trigger. For this purpose, predefined observation sequences have been developed for all  \textit{SVOM} instruments.

The Target of Opportunity Program corresponds to ground-commanded observations with short reaction times (ranging from a few minutes to a few hours). This program enables the observation of transient sources detected by other ground-based or space observatories.

Finally, the General Program covers all observations conducted by the mission in the absence of a trigger or a target of opportunity. This program is defined in advance by the Co-Investigators (Co-Is) and associated scientists. It is uploaded to the satellite by the Mission Center on a weekly basis.
The Figure~\ref{fig:map_exp} represents the fields observed by ECLAIRs from October 2024 to August 2025. There is a significant similarity with the theoretical B1 law (see Figure~\ref{fig:B1_law}), which shows that the predefined pointing strategy was well respected. Figure ~\ref{fig:map_pointings_too_gp} shows the location of the optical axis of the instruments over the same period (roughly the field of view of the VT and MXT instruments), with in blue the General Program targets and in green the targets of opportunity. We can see that the Target of Opportunity program allowed for excursions across the entire celestial sphere. 

Implementing these various scientific programs required the development of complex and highly sophisticated modules, both on board the satellite and within the ground segment, to provide greater operational flexibility. For example, enabling the platform to autonomously modify its observation schedule without relying on ground-based programming demanded dedicated developments and extensive testing \citep{Chen+etal+2026}. Moreover, rapid ground reprogramming in response to exceptional targets of opportunity was made possible through the deployment of a ‘Real-Time Response and Collaboration’ system at the Mission Center. A detailed description of this system is presented in \cite{Bai+etal+2026}. A tool for checking scheduled observations (PPST) and completed observations (AFST) has been developed by the Mission Center and can be accessed at the following address: https://soqt.smoc.ac.cn/

Once the data has been collected at the Space Science Data Centre (SSDC), it is sent to two scientific centres, one in France (FSC) and one in China (CSC), which are responsible for developing the associated scientific products. The architecture and organisation of these different centres are detailed in \cite{Liu+etal+2026}, \cite{Louvin+etal+2026} and \cite{Huang+etal+2026}.

The scientific products of the GRB Core Program and the ToO Program are made public as soon as available. With respect to GRBs, a table compiling the detected GRBs and their associated scientific products can be accessed at the following address: https://fsc.svom.org/ifsc-tools/grb-public

The scientific products of the General program remain restricted during 1 year to the PI of each accepted proposal, after which they become public.

\subsection{The Core Program}

The \textit{SVOM} spacecraft generally follows the B1 pointing law (Figure~\ref{fig:B1_law}) while awaiting GRB detection. If ECLAIRs triggers with sufficient significance \citep{Schanne+etal+2026}, the derived source position is transmitted to the spacecraft, which, after checking feasibility, autonomously slews to the target within 2-3 minutes \citep{Li+etal+2026}, thereby enabling rapid afterglow follow-up with MXT in X-rays and VT in the optical band. Following a slew, the satellite remains pointed at the source for five orbits ($\sim$8 hours). The GRB position and main characteristics derived by ECLAIRs are also rapidly transmitted to the ground through the \textit{SVOM} VHF system. Refined localizations of the X-ray counterpart provided by MXT are also downlinked via VHF. The signal is received by one of $\sim$50 ground stations distributed along the satellite ground track. The system design ensures that 63\% of alert packets reach the Science Centers within 30~s, from where they are disseminated via notices to the \textit{SVOM} ground facilities (GWAC and GFTs) and to the subscribed scientific community.

A team of scientists on call, known as the ‘Burst Advocates’ (inspired by \textit{Swift}), is on duty 24 hours a day. These scientists are responsible for responding promptly to \textit{SVOM} alerts, validating them by drafting a General Coordinates Network circular (GCN) and organising the GRB follow-up observations by ground-based facilities and/or other space observatories. To carry out these tasks, they use specific tools that have been developed and integrated within the French and Chinese scientific centres (FSC and CSC), see  \cite{Claret+etal+2026} and  \cite{Han+etal+2026}. 
 
\subsubsection{The onboard GRB observation sequence}
The GRB localization sequence represents a sophisticated orchestration of onboard instruments and systems designed to rapidly detect, localize, and observe gamma-ray bursts. This process begins with initial detection by ECLAIRs and progresses through several phases of refinement, culminating in precise localization and targeted observation.

\textbf{\begin{figure*}
    \centering
    \includegraphics[width=\textwidth, angle=0]{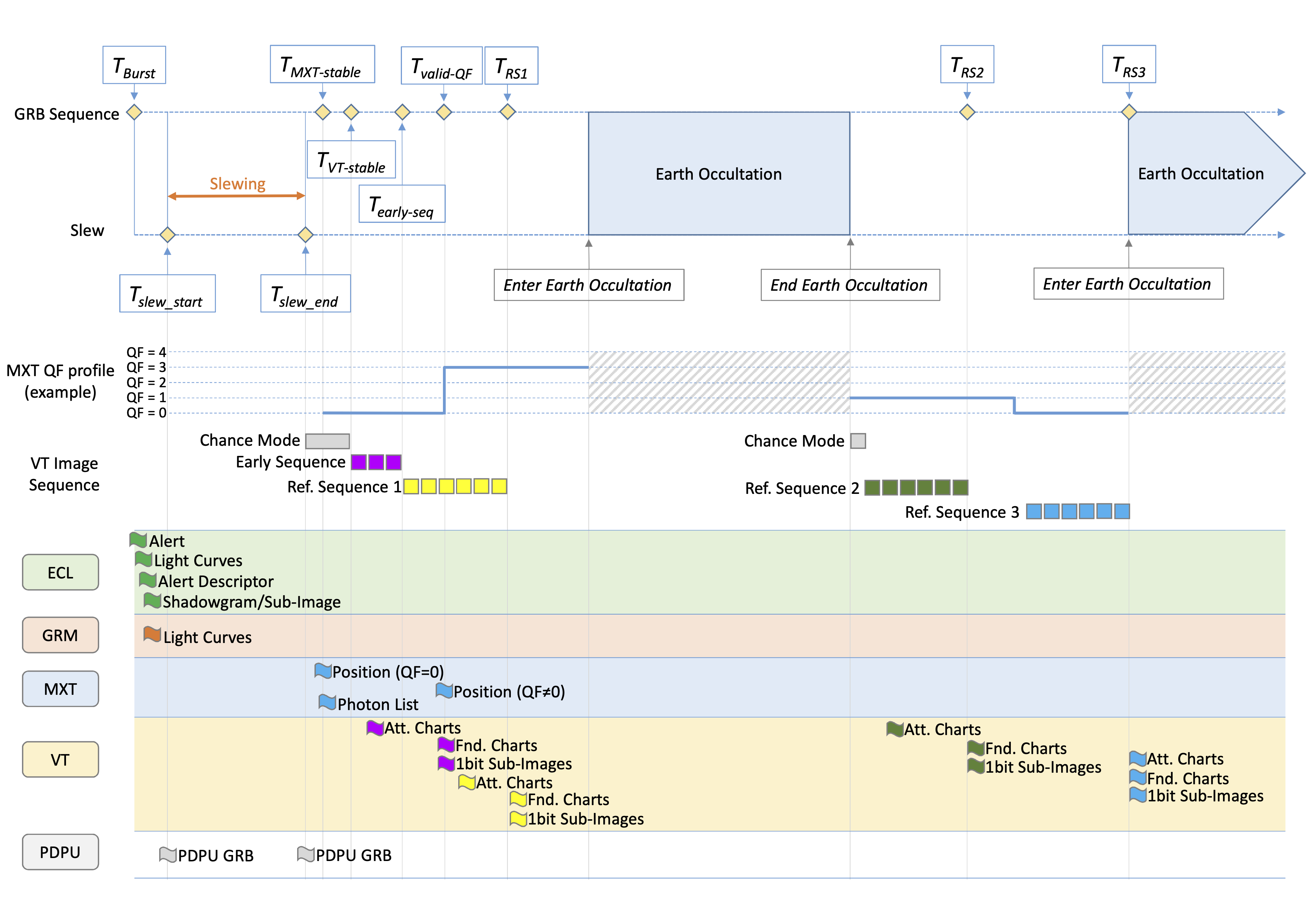}
   \caption{Schema illustrating timeline of the GRB onboard sequence, along with Quality Factor(QF) evolution (example with QF$_{max} = 3$), VT image sequences and VHF packets emission.}
   \label{fig:GRB_sequence}
\end{figure*}}

\subsubsection*{\textit{Initial Detection and Alert Generation}}

The sequence starts when ECLAIRs detects a GRB with statistical significance exceeding the predefined \textit{AlertThreshold}. Upon detection, ECLAIRs immediately generates an alert message containing crucial information about the event. This message includes the GRB observation identifier and its position as determined in ECLAIRs' reference frame \citep{Schanne+etal+2026}.

The alert message is transmitted to the Payload Data Processing Unit (PDPU) at a maximum rate of once every two seconds, continuing until either the satellite accepts a slewing maneuver to point toward the GRB or a timeout period of eight minutes elapses. 

The PDPU serves as the central coordinator of this process, performing several critical functions.
First, it relays the alert message to ground stations via the VHF link. Simultaneously, the PDPU forwards the alert to the Gamma-Ray Monitor (GRM) instrument, initiating its acquisition sequence.
In parallel, ECLAIRs begins transmitting additional scientific data to the PDPU for subsequent VHF transmission to ground. This data includes light curves that describe the intensity of the GRB over time, alert descriptors that provide detailed information about the detection parameters, and either shadowgrams or sub-images that capture the localization of the gamma-ray emission.
The GRM instrument also contributes to this data stream by transmitting its own light curve measurements to the PDPU.

\subsubsection*{\textit{Slew Request and Execution}}

A particularly significant aspect of the alert message is the potential inclusion of a slew request. ECLAIRs generates this request when it detects an unknown excess of gamma-ray emission in an image with significance greater than the \textit{SlewThreshold}. 

When the slew request is received by the PDPU, it evaluates the target attitude constraints, particularly the positions of the Sun and Moon relative to the proposed observation. If these celestial bodies would not interfere with the observation (i.e., the avoidance angles are compliant), the PDPU formulates a slew command containing the GRB quaternion as the target pointing direction, referenced to the J2000 coordinate system.

This command is transmitted to the platform, which performs its own validation checks. In particular, the platform verifies that no other slewing maneuver is currently in progress, confirms that power conditions are adequate for the maneuver, and checks the feasibility of the proposed attitude.

Assuming all validation criteria are satisfied, the platform initiates the slewing maneuver toward the specified J2000 attitude. As the maneuver begins, the platform notifies the PDPU of this status change. The PDPU, in turn, updates the Attitude and Angular Velocity (AAV) packets with the current slew status. 

As the slewing maneuver starts, ECLAIRs temporarily disables its trigger process until the maneuver is complete. PDPU generates and broadcast a PDPU GRB VHF packet, containing the summary of essential computations and checks about the targeted burst.

\subsubsection*{\textit{Early Localization Phase}}

The early localization phase represents a critical activity following GRB detection, characterized by rapid instruments responses and interactions.

Upon completion of the slewing maneuver, ECLAIRs re-enables its trigger process and, as soon as the angle deviation with the target is less than 26 arcmin, VT instrument enters a special operating mode known as \textit{chance mode}, characterized by 15-second exposure times.
MXT instrument starts its GRB localization process and this process involves several key activities. First, MXT generates position packets that contain its best estimate of the GRB location. These packets are transmitted to the PDPU at regular intervals. The localization data provided in these packets is referenced to MXT own coordinate system. In addition to localization data, MXT transmits photon lists to the PDPU, containing the list of photons detected by MXT since the GRB alert and their position on the CCD and energy level. A second PDPU GRB VHF packet is generated and broadcasted at the end of slew.

MXT stability is determined autonomously based on its own internal criteria. The moment when MXT considers its observations to be stable is designated as T$_{MXT-stable}$.
Similarly, the PDPU monitors the satellite's attitude and the AAV bulletin to determine when VT achieves stability. This determination is designated as T$_{VT-stable}$.

When VT stability is achieved, VT transitions from chance mode to early localization mode, characterized by longer exposure times and the capture of whole images covering its entire field of view. The PDPU responds to this transition by generating two attitude charts (one for the red band and the other for the blue band), identifying the brightest objects in the VT's field of view.

The sequence reaches another critical point when a valid localization from MXT becomes available. This occurs when MXT's quality factor (QF) reaches a value of 1, 2, 3, or 4: the higher the quality factor value, the greater the confidence in the localization. At this point, designated as T$_{valid-QF}$,  MXT position is used to define the location of a sub-window within the VT's whole images. PDPU will continue to receive MXT position packets, comparing their quality factors and maintaining a record of the message with the highest QF value.
VT completes its early localization sequence when it has captured three whole images. This moment is designated T$_{early-seq}$.
At the time defined as the later of T$_{valid-QF}$ and T$_{early-seq}$, PDPU generates two sets of products to be sent to ground via VHF:
\begin{itemize}
    \item Finding charts (both for the red band and blue band) that show the brightest excesses of the sources in the VT's field of view.
    \item Sub-images in 1-bit format - one for each band - that provide the VT images where pixels can have value of 0 or 1 in order to be compressed for VHF bandwidth.
\end{itemize}
This concludes the early localization sequence.

\subsubsection*{\textit{Refining Sequences}}

Following the early localization phase, the refinement process occurs in three distinct sequences.

The first refining sequence begins after the early localization sequence is complete and when the satellite achieves a stability with a drift less than 2 arcseconds over 15 seconds. During this sequence, VT captures six additional long exposure images.
Based on the images captured during this sequence, VT creates new attitude charts, finding charts and 1-bit sub-images (one product for each band). 

If a valid MXT position (QF $\neq 0 $) has been recoded by PDPU, a second slew is initiated when the satellite enters its first Earth occultation (EO). This maneuver is designed to adjust the satellite's attitude so that the GRB is centered in VT's field of view. A PDPU GRB VHF packet is generated by PDPU and broadcast via VHF link.

The second refining sequence occurs after the satellite exits its first EO and achieves the required stability threshold. During this sequence, VT captures another six long exposure images. The PDPU processes these images to generate the same set of products as in the first refining sequence.

The third and final refining sequence begins before the satellite enters its second EO. Similarly, another set of six long exposure images are taken by VT and PDPU generates the same set of products as in previous refining sequences.

\subsubsection*{\textit{GRM only sequence}}

In the event of a burst, the GRM can operate in two different modes. The instrument can either be triggered by ECLAIRs, if the latter triggers first, or it can self-trigger.

In the first case, as described in the previous section, GRM operates in a slave mode with respect to ECLAIRs and transmits a series of light curves via the VHF channel, referenced to the alert time.

In the second case, GRM enters burst mode based on its own trigger and transmits alerts, spectra, and light curves through the VHF channel. Because the GRM has a significantly larger field of view than ECLAIRs, a large number of bursts are detected by GRM alone. Since GRM does not provide accurate localization, these GRM-only alerts do not initiate an instrumental response sequence involving the satellite, MXT, or VT.

\textbf{\begin{figure*}
    \centering
    \includegraphics[width=\textwidth, angle=0]{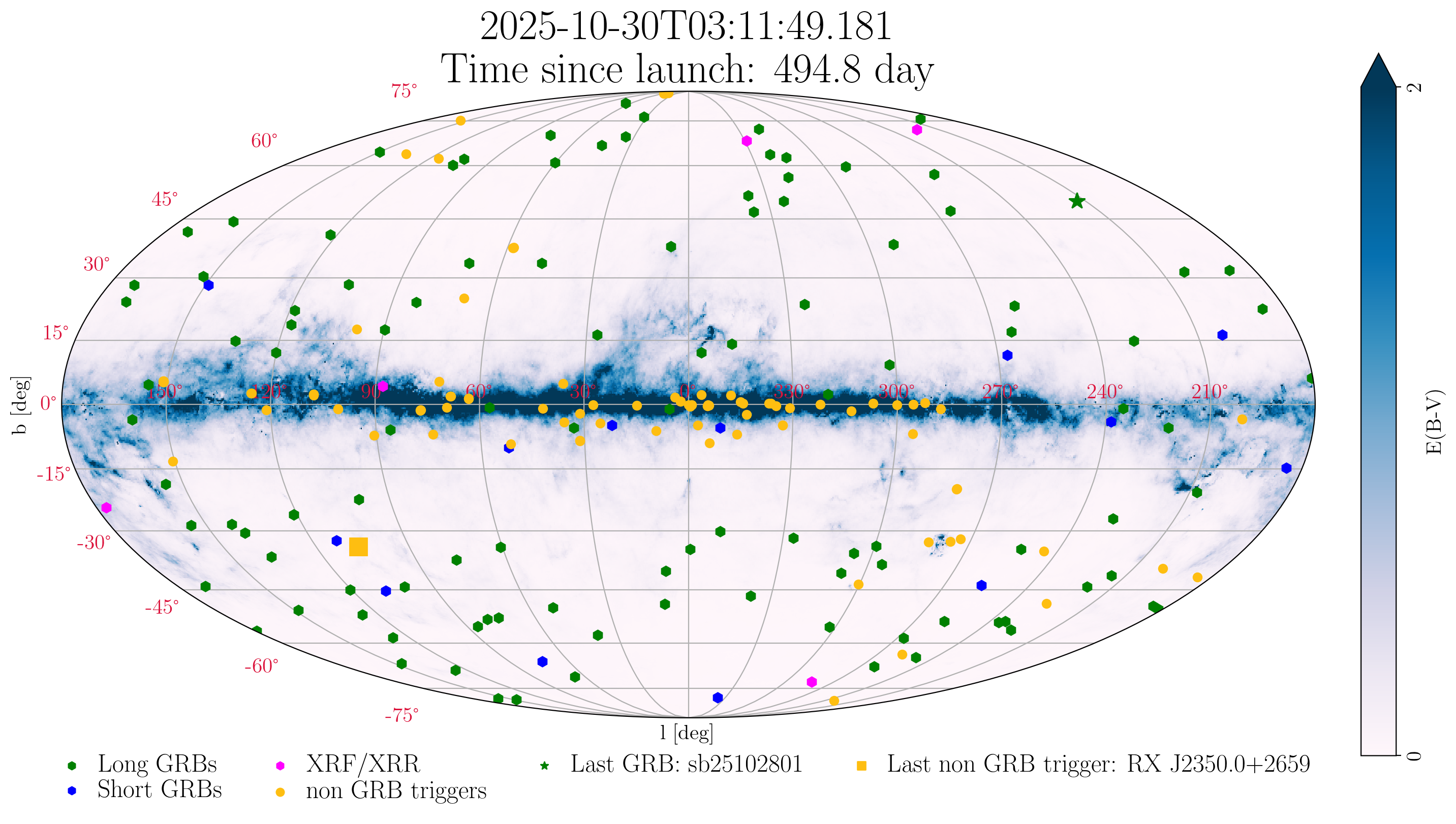}
   \caption{All \textit{SVOM} triggers (GRM + ECLAIRs) located with a localization better than 2°, since the activation of the triggers (GRM 2024 june 26, ECLAIRs 2024 July 13). This map shows the validated triggers associated with long GRBs, short GRBs and X-ray Flashes while the orange dots are the non GRB astrophysical triggers (mostly galactic x-ray binaries).
}
   \label{fig:map_GRB}
\end{figure*}}

\subsubsection{Assessment after one year}

Following the first year of operations of the \textit{SVOM} mission, i.e. at the end of October 2025, 191 triggers were confirmed as genuine Gamma-ray Bursts. Among those GRBs, 68 were detected by ECLAIRs and 165 by GRM with 40 joint detections, see Figure ~\ref{fig:map_GRB}. 
In terms of GRB detection rates, the ECLAIRs and GRM instruments are compliant with the mission requirements, which specify the localization of 200 bursts over the nominal mission period, which is expected to last three years. For ECLAIRs localized bursts, 26/68 (38\%) have a redshift measurement for the whole period of time since launch including the commissioning phase. Since the scientific exploitation phase in April 2025, this fraction of ECLAIRs GRBs with a redshift measurement has even increased up to 45\% which is also close to the mission requirements.
A rough classification of the GRB types (merger-type I vs collapsars-type II) gives us the following preliminary statistics; 33/191 (17\%) of type I bursts and 158/191 (83\%) of type II bursts.
For a more detailed description of the preliminary scientific results on gamma-ray burst science, see \cite {Daigne+etal+2026}.

\subsection{The Target of Opportunity Program}

The Target of Opportunity (ToO) program of \textit{SVOM} manages unplanned observations of transient and variable sources programmed from the ground.
All scientists have the opportunity to apply for ToOs, evaluated by the PIs.
Accepted observations are performed 
within some delay by the \textit{SVOM} space instruments, depending on the availability of the satellite up-link stations and the Beidou inter-satellite messaging system.
At system level, the ToO system was designed to perform observations within 48 h for a standard ToO, and within 12 h for an exceptional ToO (e.g. galactic supernova or Gravitational Wave alert) using, on request, additional stations. 
Multi-messenger ToOs were designed to search for counterparts of poorly localized events, such as gravitational-wave or high-energy neutrino alerts. This system enables the scheduling of optimized tiling observations for the narrow field-of-view instruments MXT and VT.
The typical observation duration was planned to be 1 orbit (45 min useful time) for a standard ToO, and 15 orbits ($\sim$24\,h) for an exceptional ToO. Multi-messenger ToOs allow up to five pointings per orbit, yielding a maximum coverage of 75 tiles over 15 orbits.
A significant fraction of the useful time, 15\% at the beginning of the mission, was foreseen for ToO observations.

\textbf{\begin{figure*}
    \centering
    \includegraphics[width=\textwidth, angle=0]{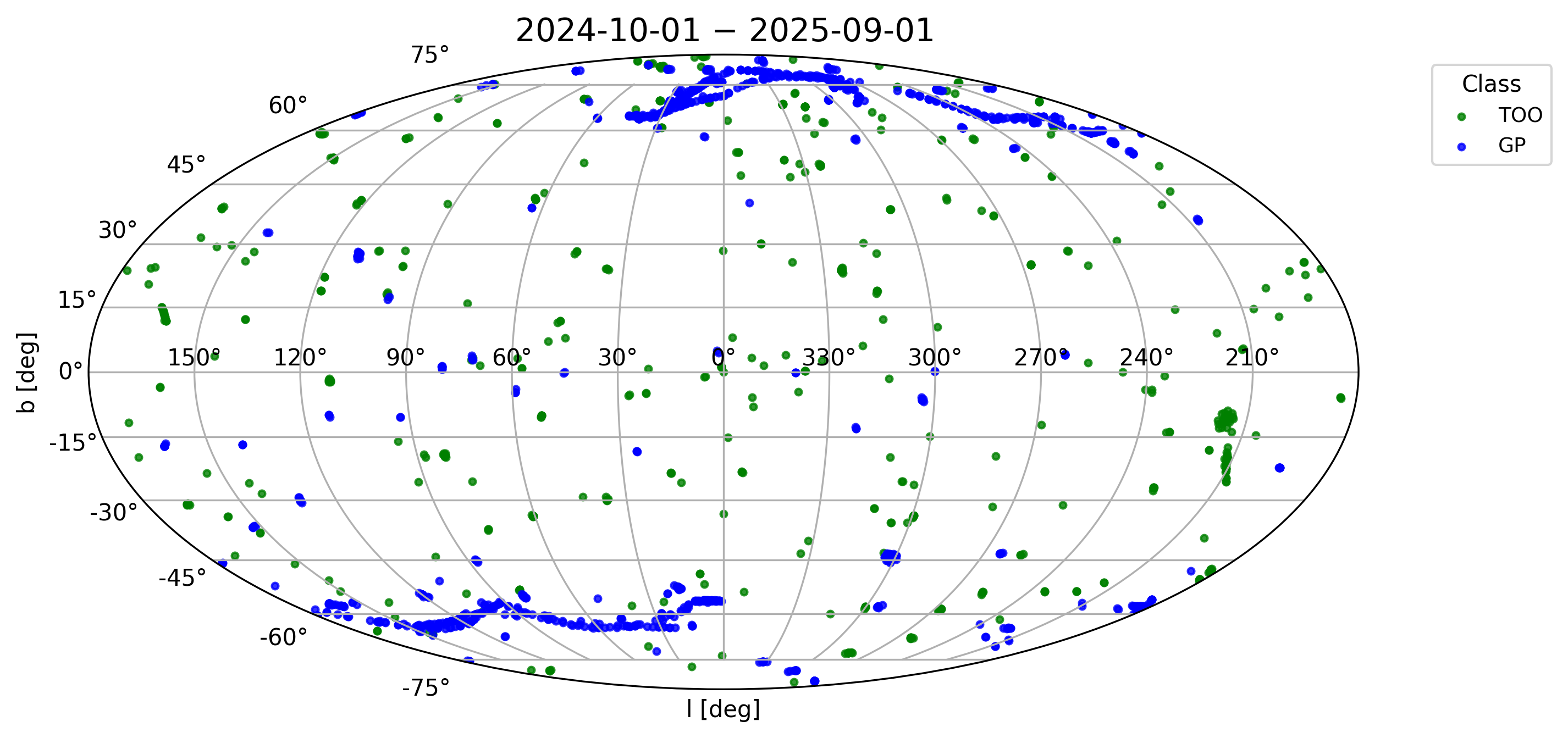}
   \caption{Distribution of the pointing directions in the sky for GP and ToO observations during almost one year, between 2024/10/01 and 2025/09/01. The lines visible near the Galactic poles correspond to the sequence of pointing that follow the B1 attitude law.}
   \label{fig:map_pointings_too_gp}
\end{figure*}}

\subsubsection{Assessment after one year}

Since 2025 January 1, an average of 14.9 targets of opportunity (ToOs) have been executed each week. These are distributed as follows: 9.4 ToOs for gamma-ray burst (GRB) follow-up, 3.4 ToOs for astrophysical targets (AT), and 2.1 ToOs for calibration purposes (CALIB).The CALIB observations are dedicated to calibration sources, such as the Crab Nebula, and may be coordinated with other missions. Each ToO typically requires an average of 2.3 orbits (similar for GRB, AT, and CALIB) resulting in a total observing load of approximately 28 orbits per week.

Non-GRB ToO activities primarily focused on monitoring outbursts from X-ray binaries and blazars, as well as following up events reported by external facilities such as Einstein Probe and KM3NeT (see \cite{Coleiro+etal+2026}).

The number of ToOs executed shows a significant increase compared to the initial forecasts, with 14.9 ToOs performed per week instead of the expected 7, and a total of 34.5 orbits per week instead of the predicted 7. Two main factors account for the significant increase in ToO activity compared to initial expectations. First, the use of the Beidou system allows satellite scheduling within a few minutes, even for standard ToOs, which is a major advantage for GRB follow-up, enabling real-time decisions on whether to continue observations. Second, the VT instrument provides highly detailed GRB monitoring without the usual ground-based limitations such as night, weather, or moonlight.

\subsection{The General Program}

The General Program (GP) of \textit{SVOM} is devoted to pre-planned observations and complements the transient sky observations managed by the Core Program and ToOs.
The GP is built taking into account the system requirements of the GRB program, in particular the pointing strategy that optimizes the ground follow-up of \textit{SVOM} GRBs. 
Therefore the GP allows pointing at sources close (i.e. within $\pm$10$^\circ$) to the B1 attitude law.
In order to increase the scientific interest of the GP, it is foreseen to allow observations outside the B1 constraint during 15\% of the GP useful time.

The minimal duration of a GP observation is 1 orbit ($\sim$40 min of effective observing time).
During its first two years of operation, GP observations is open to all \textit{SVOM} co-Investigators and affiliated scientists, who may submit proposals in response to an annual \textit{SVOM} call. These proposals are evaluated by a Time Allocation Committee based on scientific merit. Starting in 2027, a guest observer program will be launched, inviting proposals for GP observations from the general scientific community. 

To support the growth of time-domain astronomy in the coming decade, an evolution of the \textit{SVOM} time allocation is planned for the extended mission phase. The fraction of time reserved for Targets of Opportunity (ToOs) will increase from 15\% to 40\%, allowing for up to five ToOs per day instead of just one or two. At the same time, the fraction of GP time available outside the B1 pointing constraint will rise from 10\% to 50\%.  

\subsubsection{Assessment after one year}

Following the first year of operations of the \textit{SVOM} mission, the GP has demonstrated overall efficiency both in terms of coverage and data quality. Out of the 1,288 sources initially planned, corresponding to an effective observation time of 3.4~Ms, 850 sources were successfully observed, yielding 2.6~Ms of usable exposure with the VT telescope. This corresponds to a completion rate of $\sim$76\%. 

In terms of planned observation time, the GP pointings represent 33.3\% of the total, a contribution slightly lower than that of standard B1 pointing (41.8\%). In terms of executed observations, this proportion remained stable, with the GP accounting for 34.9\% of the total usable exposure time (32.6\% for B1 pointings). This consistency highlights the robustness of the scheduling strategy.

Within this framework, the GP emerges as a central pillar of \textit{SVOM}’s scientific strategy, providing a good coverage of sources and contributing to the mission objectives. 
For a more detailed description of the scientific results on Observatory Science, see \cite{Coleiro+etal+2026}.

\section{ Conclusion: \textit{SVOM} in the Astronomical Panorama}
\label{sub:panorama}

The astronomical panorama of the coming decade will be shaped by new instruments designed to address many of the outstanding questions in contemporary astrophysics. This panorama includes large radio, infrared, optical, and gamma-ray telescopes, advanced gravitational wave interferometers, and cubic-kilometer class neutrino detectors, as well as simulations enabled by high-performance computing.

Within this panorama, \textit{SVOM} brings two specific capabilities: monitoring of the hard X-ray sky for high-energy transients, and multi-wavelength follow-up observations of both self-triggered high-energy transients and outstanding transients detected by other facilities through ToO observations.

\textit{SVOM} provides GRB and high-energy transient alerts at the expected rate of about 60/yr for ECLAIRs and two times more for GRM. It expands the realm of detected transients to soft events like X-ray Flashes or supernova shock breakout. Combined with other high-energy satellites it contributes to increase the probability of joint detections with gravitational wave interferometers, a crucial goal of gravitational wave astronomy, as demonstrated by the simultaneous detection of GRB 170817A and GW170817 about 8 years ago. 

Focusing on GRBs, the joint operation of \textit{SVOM} with \textit{Swift} and \textit{Einstein Probe} (\textit{EP}, \citealt{Yuan+etal+2025}) is especially fruitful to pinpoint the X-ray afterglows of \textit{SVOM} GRBs with \textit{Swift}/XRT and \textit{EP}/FXT and in addition to detect the visible afterglows of \textit{Swift}/BAT and \textit{EP}/WXT GRBs with \textit{SVOM}/VT. The sensitivity to low energy transients leads to the frequent detection of X-Ray Flashes and soft GRBs. ECLAIRs thus fills the gap between \textit{Swift}/BAT and \textit{Fermi}/GBM \citep{Meegan+etal+2009} on one side, and \textit{EP}/WXM on the other. Together these instruments permit the detection and localization of high-energy transients from sub-keV energies to hundreds of keV. This coverage provides an unprecedented view of the zoology of high-energy transients.

The synergy between these high-energy satellites, combined with the excellent follow-up of high-energy transients by large ground-based telescopes will make possible to obtain a large number of redshift: for bursts located by ECLAIRs, $\sim$40\% have a measured spectroscopic redshift, $\sim$25\% have an optical afterglow with an accuracy of a few arcseconds, which should make it possible to measure a redshift a posteriori. Finally, the remaining $\sim$35\% without an optical afterglow includes high redshift GRB candidates ($\sim$5\% of ECLAIRs GRBs) for which we have an X-ray localization and early optical deep limits with VT. Indeed, in this context, an interesting feature of \textit{SVOM} is its capability to detect and quickly recognize very distant GRBs, like GRB 250314A at redshift z = 7.3 \citep{Cordier+etal+2025}, thanks to their soft spectrum and lack of optical emission in the VT, down to deep limits, few minutes after the burst.

After a successful launch in June 2024, and more than 1 year of flawless operation, \textit{SVOM} has taken its full place in the panoply of missions dedicated to the detection and observation of explosive transients.

This volume provides a comprehensive overview of all aspects of the mission and is intended as a resource for researchers planning to work with its data, as well as for readers interested in the mission’s technical and scientific details.

Readers interested in following the \textit{SVOM} mission are invited to explore the latest scientific highlights and mission updates at the following address: https://www.svom.eu/en/home/

\begin{acknowledgements}
The Space-based multi-band astronomical Variable Objects Monitor (\textit{SVOM}) is a joint Chinese-French mission led by the Chinese National Space Administration (CNSA), the French Space Agency (CNES), and the Chinese Academy of Sciences (CAS). We gratefully acknowledge the unwavering support of NSSC, IAMCAS, XIOPM, NAOC, IHEP, CNES, CEA, and CNRS.

The SVOM team would like to express their deep gratitude to Jacques Paul, Pierre Mandrou, Hu Jingyao, and Li Shouchen. Without their significant contributions, the SVOM mission would not have been possible. 
\end{acknowledgements}

\label{lastpage}
\bibliography{bibtex}

@INPROCEEDINGS{Cordier+etal+2008,
       author = {{Cordier}, B. and {Desclaux}, F. and {Foliard}, J. and {Schanne}, S.},
        title = "{SVOM pointing strategy: how to optimize the redshift measurements?}",
    booktitle = {Gamma-ray Bursts 2007},
         year = 2008,
       editor = {{Galassi}, M. and {Palmer}, David and {Fenimore}, Ed},
       series = {American Institute of Physics Conference Series},
       volume = {1000},
        month = may,
    publisher = {AIP},
        pages = {585-588},
          doi = {10.1063/1.2943538},
archivePrefix = {arXiv},
       eprint = {0807.0739},
 primaryClass = {astro-ph},
       adsurl = {https://ui.adsabs.harvard.edu/abs/2008AIPC.1000..585C}
}

@ARTICLE{Cordier+etal+2025,
       author = {{Cordier}, B. and {Wei}, J.~Y. and {Tanvir}, N.~R. and {Vergani}, S.~D. and {Malesani}, D.~B. and {Fynbo}, J.~P.~U. and {de Ugarte Postigo}, A. and {Saccardi}, A. and {Daigne}, F. and {Atteia}, J.-L. and {Godet}, O. and {G{\"o}tz}, D. and {Qiu}, Y.~L. and {Schanne}, S. and {Xin}, L.~P. and {Zhang}, B. and {Zhang}, S.~N. and {Nayana}, A.~J. and {Piro}, L. and {Fausey}, H. and {Schneider}, B. and {Levan}, A.~J. and {Thakur}, A.~L. and {Zhu}, Z.~P. and {Corcoran}, G. and {Rakotondrainibe}, N.~A. and {D'Elia}, V. and {Turpin}, D. and {Ag{\"u}{\'\i} Fern{\'a}ndez}, J.~F. and {Aloy}, M.~A. and {An}, J. and {Bai}, M. and {Basa}, S. and {Bernardini}, M.~G. and {Bochenek}, A. and {Brivio}, R. and {Brunet}, M. and {Bruni}, G. and {Cenko}, S.~B. and {Cheng}, Q. and {Chrimes}, A. and {Christensen}, L. and {Claret}, A. and {Coleiro}, A. and {Cotter}, L. and {Crepaldi}, S. and {Deng}, J.~S. and {Dimple} and {Dong}, Y.~W. and {Dornic}, D. and {Evans}, P.~A. and {Eyles-Ferris}, R.~A.~J. and {Ferro}, M. and {Galbany}, L. and {Garnichey}, M. and {Gianfagna}, G. and {Gompertz}, B.~P. and {Goto}, H. and {Habeeb}, N. and {Han}, P.~Y. and {Han}, X.~H. and {Hartmann}, D.~H. and {Heintz}, K.~E. and {Hu}, J.~Y. and {Huang}, M.~H. and {Izzo}, L. and {Jakobsson}, P. and {Kennea}, J.~A. and {Lachaud}, C. and {Laskar}, T. and {Li}, D. and {Li}, H.~L. and {Li}, R.~Z. and {Liu}, X. and {Liu}, Y. and {Lombardi}, G. and {Louvin}, H. and {Maggi}, P. and {Maiolino}, T. and {Mao}, Q.~Y. and {Martin-Carrillo}, A. and {Mercier}, K. and {O'Brien}, P. and {Palmerio}, J.~T. and {Petitjean}, P. and {Pieterse}, D.~L.~A. and {Piron}, F. and {Pugliese}, G. and {Rayson}, B.~C. and {Reynolds}, T. and {Robinet}, F. and {Rossi}, A. and {Salvaterra}, R. and {Th{\"o}ne}, C.~C. and {Top{\c{c}}u}, B. and {Wang}, C.~W. and {Wang}, J. and {Wang}, Y. and {Wu}, C. and {Xiong}, S.~L. and {Xu}, D. and {Yang}, H.~N. and {Yuan}, W.~M. and {Zhang}, Y.~H. and {Zhang}, X.~F. and {Zheng}, S.~J.},
        title = "{SVOM GRB 250314A at z ≃ 7.3: An exploding star in the era of re-ionization}",
      journal = {\aap},
         year = 2025,
        month = dec,
       volume = {704},
          eid = {L7},
        pages = {L7},
          doi = {10.1051/0004-6361/202556580},
archivePrefix = {arXiv},
       eprint = {2507.18783},
 primaryClass = {astro-ph.HE},
       adsurl = {https://ui.adsabs.harvard.edu/abs/2025A&A...704L...7C}
}

@ARTICLE{Cordier+etal+2026b,
   author = {{Cordier}, B. and {Jeannin}, L. and {Lafabrie}, Ph. and {Chavanas}, G. and {Crepaldi}, S. and {Dagoneau}, N. and {Formica}, A. and {Garcia}, V. and {Jolivet}, L. and {Lacour}, S. and {Louvin}, H. and {Sabatier}, E. },
    title = "{The VHF alert network of the SVOM mission}",
    journal = {\raa},
    year = 2026,
    volume = {this issue},
    pages = {1-10},
}

@ARTICLE{Gehrels+etal+2004,
       author = {{Gehrels}, N. and {Chincarini}, G. and {Giommi}, P. and {Mason}, K.~O. and {Nousek}, J.~A. and {Wells}, A.~A. and {White}, N.~E. and {Barthelmy}, S.~D. and {Burrows}, D.~N. and {Cominsky}, L.~R. and {Hurley}, K.~C. and {Marshall}, F.~E. and {M{\'e}sz{\'a}ros}, P. and {Roming}, P.~W.~A. and {Angelini}, L. and {Barbier}, L.~M. and {Belloni}, T. and {Campana}, S. and {Caraveo}, P.~A. and {Chester}, M.~M. and {Citterio}, O. and {Cline}, T.~L. and {Cropper}, M.~S. and {Cummings}, J.~R. and {Dean}, A.~J. and {Feigelson}, E.~D. and {Fenimore}, E.~E. and {Frail}, D.~A. and {Fruchter}, A.~S. and {Garmire}, G.~P. and {Gendreau}, K. and {Ghisellini}, G. and {Greiner}, J. and {Hill}, J.~E. and {Hunsberger}, S.~D. and {Krimm}, H.~A. and {Kulkarni}, S.~R. and {Kumar}, P. and {Lebrun}, F. and {Lloyd-Ronning}, N.~M. and {Markwardt}, C.~B. and {Mattson}, B.~J. and {Mushotzky}, R.~F. and {Norris}, J.~P. and {Osborne}, J. and {Paczynski}, B. and {Palmer}, D.~M. and {Park}, H.-S. and {Parsons}, A.~M. and {Paul}, J. and {Rees}, M.~J. and {Reynolds}, C.~S. and {Rhoads}, J.~E. and {Sasseen}, T.~P. and {Schaefer}, B.~E. and {Short}, A.~T. and {Smale}, A.~P. and {Smith}, I.~A. and {Stella}, L. and {Tagliaferri}, G. and {Takahashi}, T. and {Tashiro}, M. and {Townsley}, L.~K. and {Tueller}, J. and {Turner}, M.~J.~L. and {Vietri}, M. and {Voges}, W. and {Ward}, M.~J. and {Willingale}, R. and {Zerbi}, F.~M. and {Zhang}, W.~W.},
        title = "{The Swift Gamma-Ray Burst Mission}",
      journal = {\apj},
         year = 2004,
        month = aug,
       volume = {611},
       number = {2},
        pages = {1005-1020},
          doi = {10.1086/422091},
archivePrefix = {arXiv},
       eprint = {astro-ph/0405233},
 primaryClass = {astro-ph},
       adsurl = {https://ui.adsabs.harvard.edu/abs/2004ApJ...611.1005G}
}

@ARTICLE{Godet+etal+2026a,
   author = {{Godet}, O. and {Atteia}, J.~L. and {Schanne}, S.  and 
	{Lachaud}, C. and {Goldwurm}, A. and {others} },
    title = "{ECLAIRs:the SVOM high-energy transient trigger camera}",
  journal = {\raa},
   year = 2026,
   volume = {this issue},
   pages = {1-10},
}

@ARTICLE{Goetz+etal+2026,
   author = {{G\"otz}, D. and {Crepaldi}, S. and {Doumayrou}, E. and {Feldman}, C. and {Ferrando}, P. and {Fort}, A. },
   title = "The Microchannel X-ray Telescope on board the SVOM mission: in flight
scientific performance",
   journal = {\raa},
   year = 2026,
   volume = {this issue},
    pages = {1-10},
}

@ARTICLE{Li+etal+2026,
   author = {{Li}, D. and {Zhang}, Y. and {Su}, R. and {Qin}, G. and {Zhang}, X. and {Chen}, W.},
   title = "{An Overview of SVOM Satellite’s Design and In-orbit Operations}",
   journal = {\raa},
   year = 2026,
   volume = {this issue},
   pages = {1-10},
}

@ARTICLE{Qiu+etal+2026,
   author = {{Qiu}, Y.~L. and {Wang}, J.~M. and {Ho}, L.~C. and {Chen}, Y.-M. and {Zhang}, H.~T. and {Bian}, W.~H. and {Xue}, S.~J. and {others}},
   title = "{An Overview of the Visible Telescope on Board the SVOM Mission}",
   journal = {\raa},
   year = 2026,
   volume = {this issue},
   pages = {1-10},
}

@ARTICLE{Sun+etal+2026,
   author = {{Sun}, J.~C. and {Dong}, Y.~W. and {He}, J. and {Liu}, J.~T. and {Li}, L. and {Wang}, R.~J.},
   title = "{The Gamma-Ray Monitor onboard the SVOM satellite}",
   journal = {\raa},
   year = 2026,
   volume = {this issue},
   pages = {1-10},
}

@ARTICLE{Bai+etal+2026,
   author = {{Bai}, M. and {SU}, J. and {Li}, B. and {Feng}, Z. and {Man}, Y. and {Xiao}, Z. and {Liu}, Y. and {HU}, T.},
    title = "{SVOM Real-time Response and Collaboration System}",
    journal = {\raa},
    year = 2026,
    volume = {this issue},
    pages = {1-10},
}

@ARTICLE{Liu+etal+2026,
   author = {{Liu}, Y. and {Bai}, M. and {Wei}, M. and {Feng}, K. and{Li}, B.},
   title = "{SVOM Ground Support System}",
   journal = {\raa},
    year = 2026,
   volume = {this issue},
    pages = {1-10},
}

@ARTICLE{Xin+etal+2026,
   author = {{Xin}, L. and {Huang}, L. and {Cai}, H. and {Han}, X. and {Xu}, Y. and {Lu}, X.},
    title = "{Overview of Ground-based WideAngle Camera array}",
    journal = {\raa},
    year = 2026,
   volume = {this issue},
    pages = {1-10},
}

@ARTICLE{Wu+etal+2026,
   author = {{Wu}, C. and {Kang}, Z. and {Lu}, X.~M. and {Han}, X. and {Xin}, L. and {Zhang}, P.},
    title = "{SVOM/C-GFT : Instrumentation and Early Science Performances}",
    journal = {\raa},
    year = 2026,
   volume = {this issue},
    pages = {1-10},
}

@ARTICLE{Basa+etal+2026,
   author = {{Basa}, S. and {Lee}, W.H. and {Watson}, A.M. and {Dolon}, F. and {floriot}, J. and {Atteia}, J.-L. and {others}},
    title = "{SVOM/FM-GFT : Instrumentation and Performances on the SVOM Alerts}",
    journal = {\raa},
    year = 2026,
   volume = {this issue},
    pages = {1-10},
}

@ARTICLE{Daigne+etal+2026,
   author = {{Daigne}, F. and {Turpin}, D. and {Atteia}, J.-L. and {Palmerio}, J. and {Xin}, L. and {Vergani}, S. and {others}},
    title = "{First Gamma-Ray Burst Observations with SVOM}",
    journal = {\raa},
    year = 2026,
   volume = {this issue},
    pages = {1-10},
}

@ARTICLE{Coleiro+etal+2026,
   author = {{Coleiro}, A. and {Tao}, L. and {Cangemi}, F. and {Han}, X. and {Brunet}, M. and {Dagoneau}, N. and {others} },
    title = "{Early results from the SVOM Observatory Science program}",
    journal = {\raa},
    year = 2026,
   volume = {this issue},
    pages = {1-10},
}

@ARTICLE{Louvin+etal+2026,
   author = {{Louvin}, H. and {Corre}, D. and {Formica}, A. and {Jouvin}, L. and {Tazhenova}, K. and {Sadibekova}, T. and {others}},
    title = "{The SVOM French Science Center Infrastructure}",
    journal = {\raa},
    year = 2026,
   volume = {this issue},
    pages = {1-10},
}

@ARTICLE{Huang+etal+2026,
   author = {{Huang}, M. and {Zheng}, S.~J. and {others}},
    title = "{Chinese Science Center Infrastructure}",
    journal = {\raa},
    year = 2026,
   volume = {this issue},
    pages = {1-10},

}

@ARTICLE{Chen+etal+2026,
   author = {{Chen}, K. and {Zhang}, Y.H. and {Su}, R.~F. and {Mao}, Q.~Y. and {He}, J.~W. and {Chen}, W. and {Zhang}, X.~F. and {others} },
    title = "{The SVOM observation management strategy and in-orbit verification}",
    journal = {\raa},
    year = 2026,
   volume = {this issue},
    pages = {1-10},
}

@ARTICLE{Meegan+etal+2009,
       author = {{Meegan}, Charles and {Lichti}, Giselher and {Bhat}, P.~N. and {Bissaldi}, Elisabetta and {Briggs}, Michael S. and {Connaughton}, Valerie and {Diehl}, Roland and {Fishman}, Gerald and {Greiner}, Jochen and {Hoover}, Andrew S. and {van der Horst}, Alexander J. and {von Kienlin}, Andreas and {Kippen}, R. Marc and {Kouveliotou}, Chryssa and {McBreen}, Sheila and {Paciesas}, W.~S. and {Preece}, Robert and {Steinle}, Helmut and {Wallace}, Mark S. and {Wilson}, Robert B. and {Wilson-Hodge}, Colleen},
        title = "{The Fermi Gamma-ray Burst Monitor}",
      journal = {\apj},
         year = 2009,
        month = sep,
       volume = {702},
       number = {1},
        pages = {791-804},
          doi = {10.1088/0004-637X/702/1/791},
archivePrefix = {arXiv},
       eprint = {0908.0450},
 primaryClass = {astro-ph.IM},
       adsurl = {https://ui.adsabs.harvard.edu/abs/2009ApJ...702..791M}
}

@ARTICLE{Yuan+etal+2025,
       author = {{Yuan}, Weimin and {Dai}, Lixin and {Feng}, Hua and {Jin}, Chichuan and {Jonker}, Peter and {Kuulkers}, Erik and {Liu}, Yuan and {Nandra}, Kirpal and {O'Brien}, Paul and {Piro}, Luigi and {Rau}, Arne and {Rea}, Nanda and {Sanders}, Jeremy and {Tao}, Lian and {Wang}, Junfeng and {Wu}, Xuefeng and {Zhang}, Bing and {Zhang}, Shuangnan and {Ai}, Shunke and {Buchner}, Johannes and {Bulbul}, Esra and {Chen}, Hechao and {Chen}, Minghua and {Chen}, Yong and {Chen}, Yu-Peng and {Coleiro}, Alexis and {Coti Zelati}, Francesco and {Dai}, Zigao and {Fan}, Xilong and {Fan}, Zhou and {Friedrich}, Susanne and {Gao}, He and {Ge}, Chong and {Ge}, Mingyu and {Geng}, Jinjun and {Ghirlanda}, Giancarlo and {Gianfagna}, Giulia and {Gou}, Lijun and {Guillot}, S{\'e}bastien and {Hou}, Xian and {Hu}, Jingwei and {Huang}, Yongfeng and {Ji}, Long and {Jia}, Shumei and {Komossa}, S. and {Kong}, Albert K.~H. and {Lan}, Lin and {Li}, An and {Li}, Ang and {Li}, Chengkui and {Li}, Dongyue and {Li}, Jian and {Li}, Zhaosheng and {Ling}, Zhixing and {Liu}, Ang and {Liu}, Jinzhong and {Liu}, Liangduan and {Liu}, Zhu and {Luo}, Jiawei and {Ma}, Ruican and {Maggi}, Pierre and {Maitra}, Chandreyee and {Marino}, Alessio and {Ng}, Stephen Chi-Yung and {Pan}, Haiwu and {Rukdee}, Surangkhana and {Soria}, Roberto and {Sun}, Hui and {Tam}, Pak-Hin Thomas and {Thakur}, Aishwarya Linesh and {Tian}, Hui and {Troja}, Eleonora and {Wang}, Wei and {Wang}, Xiangyu and {Wang}, Yanan and {Wei}, Junjie and {Wen}, Sixiang and {Wu}, Jianfeng and {Wu}, Ting and {Xiao}, Di and {Xu}, Dong and {Xu}, Renxin and {Xu}, Yanjun and {Xu}, Yu and {Yang}, Haonan and {You}, Bei and {Yu}, Heng and {Yu}, Yunwei and {Zhang}, Binbin and {Zhang}, Chen and {Zhang}, Guobao and {Zhang}, Liang and {Zhang}, Wenda and {Zhang}, Yu and {Zhou}, Ping and {Zou}, Zecheng},
        title = "{Science objectives of the Einstein Probe mission}",
      journal = {Science China Physics, Mechanics, and Astronomy},
         year = 2025,
        month = mar,
       volume = {68},
       number = {3},
          eid = {239501},
        pages = {239501},
          doi = {10.1007/s11433-024-2600-3},
archivePrefix = {arXiv},
       eprint = {2501.07362},
 primaryClass = {astro-ph.HE},
       adsurl = {https://ui.adsabs.harvard.edu/abs/2025SCPMA..6839501Y}
}

@ARTICLE{Han+etal+2026,
   author = {{Han}, X.~H. and {Xin}, L.~P. and {Wang}, J. and {XIao}, Y.~J. and {Zhang}, P.~P. and {Cai}, H.~B. and {others}},
    title = "{SVOM Science User Support at CSC}",
    journal = {\raa},
    year = 2026,
    volume = {this issue},
    pages = {1-10},

}

@ARTICLE{Claret+etal+2026,
   author = {{Claret}, A. and {Turpin}, D. and {Moreau}, C. and {Thome}, J.-C. and {Sadibekova}, T. and {others}},
    title = "{SVOM Science User Support at FSC}",
    journal = {\raa},
    year = 2026,
   volume = {this issue},
    pages = {1-10},
}

@ARTICLE{Schanne+etal+2026,
   author = {{Schanne}, S. and {Chateau}, F. and {Dagoneau}, N. and {Le Provost}, H. and {Kestener}, P. and {others}},
    title = "{Overview of the ECLAIRs trigger for SVOM gamma-ray burst detection}",
    journal = {\raa},
    year = 2026,
   volume = {this issue},
    pages = {1-10},
}

@article{Zhang+etal+2009,
  author       = {Zhang, Bing and Zhang, Bin-Bin and Virgili, Francisco J. and Liang, En-Wei and Kann, David A. and Wu, Xue-Feng and Proga, Daniel and Lv, Hou-Jun and Toma, Ko and Mészáros, Peter and Burrows, David N. and Roming, Peter W. A. and Gehrels, Neil},
  title        = {Discerning the Physical Origins of Cosmological Gamma-ray Bursts Based on Multiple Observational Criteria: The Cases of $z=6.7$ GRB~080913, $z=8.3$ GRB~090423, and Some Short/Hard GRBs},
  journal      = {The Astrophysical Journal},
  volume       = {703},
  number       = {2},
  pages        = {1696--1724},
  year         = {2009},
  doi          = {10.1088/0004-637X/703/2/1696},
  eprint       = {0902.2419},
  archivePrefix= {arXiv},
  primaryClass = {astro-ph.HE}
}

@article{Cucchiara+etal+2011,
	Adsurl = {http://adsabs.harvard.edu/abs/2011ApJ...736....7C},
	Archiveprefix = {arXiv},
	Author = {{Cucchiara}, A. and {Levan}, A.~J. and {Fox}, D.~B. and {Tanvir}, N.~R. and {Ukwatta}, T.~N. and {Berger}, E. and {Kr{\"u}hler}, T. and {K{\"u}pc{\"u} Yolda{\c s}}, A. and {Wu}, X.~F. and {Toma}, K. and {Greiner}, J. and {Olivares}, F.~E. and {Rowlinson}, A. and {Amati}, L. and {Sakamoto}, T. and {Roth}, K. and {Stephens}, A. and {Fritz}, A. and {Fynbo}, J.~P.~U. and {Hjorth}, J. and {Malesani}, D. and {Jakobsson}, P. and {Wiersema}, K. and {O'Brien}, P.~T. and {Soderberg}, A.~M. and {Foley}, R.~J. and {Fruchter}, A.~S. and {Rhoads}, J. and {Rutledge}, R.~E. and {Schmidt}, B.~P. and {Dopita}, M.~A. and {Podsiadlowski}, P. and {Willingale}, R. and {Wolf}, C. and {Kulkarni}, S.~R. and {D'Avanzo}, P.},
	Doi = {10.1088/0004-637X/736/1/7},
	Eid = {7},
	Eprint = {1105.4915},
	Journal = {\apj},
	Month = jul,
	Pages = {7},
	Primaryclass = {astro-ph.CO},
	Title = {{A Photometric Redshift of z \~{} 9.4 for GRB 090429B}},
	Volume = 736,
	Year = 2011}

@article{ViglianoLongo+2024,
  author       = {Vigliano, Alessandro Armando and Longo, Francesco},
  title        = {Gamma-ray Bursts: 50 Years and Counting!},
  journal      = {Universe},
  volume       = {10},
  number       = {2},
  pages        = {57},
  year         = {2024},
  doi          = {10.3390/universe10020057},
  url          = {https://www.mdpi.com/2218-1997/10/2/57}
}

\end{document}